\documentclass[twocolumn,showpacs,amsmath,nofootinbib,amssymb]{revtex4}

\usepackage{graphicx}
\usepackage{bm}

\begin{document}

\bibliographystyle{apsrev}
\title{Monte Carlo Neutrino Oscillations}

\author{James P. Kneller}
\email{Kneller@physics.umn.edu}
\affiliation{Department of Physics, North Carolina State University, Raleigh, North Carolina 27695-8202}
\affiliation{School of Physics and Astronomy, University of Minnesota, Minneapolis, Minnesota 55455}

\author{Gail C. McLaughlin}
\email{Gail_McLaughlin@ncsu.edu}
\affiliation{Department of Physics, North Carolina State University, Raleigh, North Carolina 27695-8202}

\date{\today}


\begin{abstract}

We demonstrate that the effects of matter upon neutrino propagation may be recast as the scattering of the initial neutrino wavefunction. Exchanging the differential, Schrodinger equation for an integral equation for the scattering matrix $S$ permits a Monte Carlo method for the computation of $S$ that removes many of the numerical difficulties
associated with direct integration techniques. 
\end{abstract}

\maketitle

\newpage

\section{Introduction} \label{sec:intro}

As a neutrino propagates through matter the non-zero density modulates the flavor oscillations of the neutrino wavefunction. The evolution of the wavefunction differs from that in vacuum with the consequnce that neutrino flavour transformation may be enhanced. Appreciation of this effect, first discussed by Mikheyev \& Smirnov and Wolfenstein \cite{M&S1986,Wolfenstein1977}, is particularly important when the source of neutrinos is buried deep within dense matter such as those one can find in astrophysical settings. Indeed, this transformation was first invoked to resolve the discrepancy between the observed and predicted detection rates of solar electron neutrinos \cite{Bethe1986,Haxton1986} and the most compelling experimental evidence for this effect has come from the Sudbury Neutrino Observatory which is capable of 
measuring the $\mu$ or $\tau$ flavor content of the neutrinos initially produced in the center of the Sun as electron type \cite{SNO}. 
In the same fashion, the flavor content of neutrinos emitted from the neutrinosphere in a proto-neutron star will be altered by their propagation through the overlying progenitor material \cite{fuller, dighe,SF2002,EMV2003,lunardini,BBKM2005} and it is also apparent that understanding the matter effect of the Earth is crucial for interpreting any future long baseline experiment, see e.g. \cite{mocioiu}. 

For each of these situations one is provided with the initial state of the neutrino at the source and wishes to determine the flavor content of the wavefunction after it has passed through the intervening material. Gauging the matter effects means possessing a suitable calculational tool. The most obvious point of departure for such a calculation is the Schrodinger equation. Since there are three neutrino flavors the neutrino wavefunction must posses three complex components and the Hamiltonian, $H$, is a 3 $\times$ 3 matrix. In vacuum the Hamiltonian in the flavor basis is not diagonal and it is the presence of the off-diagonal terms in $H$ that lead to flavor oscillations. The vacuum Hamiltonian may be diagonalized by a suitable unitary transformation and it is this new basis that form the `mass eigenstates'. 

In the presence of matter a potential, $V(x)$, that takes into account coherent forward scattering of the neutrinos, must be included in the Hamiltonian. For the case of only active neutrino flavors (i.e. all the flavors 
that have ordinary weak interactions) passing through normal matter the only relevant portion of $V(x)$ is the $\nu_{e}-\nu_{e}$ component of $V(x)$. This is the well-known $V(x)=\sqrt{2}\,G_F \rho(x)$ where $G_F$ is Fermi's constant and $\rho(x)$ is the electron number density. With the addition of $V(x)$ the Hamiltonian is no longer diagonal in the mass basis. A new basis, the `matter eigenstates', diagonalizes $H(x)$ but the spatial variance now within the Hamiltonian means that the unitary transformtion that relates the flavor to the matter basis also varies with the propagtion distance. The gradient of this unitary transformation is non-zero and one finds that the Schrodinger equation in this new basis picks up off-diagonal terms. Again, the presence of off-diagonal terms in the Schrodinger equation leads to mixing of the complex coefficients describing the wavefunction and this will occur even in the matter basis if those terms are sufficiently large.

Though, in general, the three complex components of the wavefunction oscillate simulataneously the large difference in vacuum mass splittings usually permits us to consider the evolution of the neutrino wavefunction as being factored into two, localized, spatially separated, two-neutrino mixings. This factorization simplifies matters greatly. For two-neutrino mixing there is a single rotation angle $\theta_V$ that describes the relationship between the two mass eigenstates and the flavor states, and, similarly, within matter there is only one rotation angle $\theta(x)$, the matter mixing angle, for the relationship between the matter and flavor bases. In the matter eigenstate basis the Schrodinger equation for the evolution of the 2-component neutrino wavefunction is
\begin{equation}
\imath \frac{d}{dx} \left(\begin{array}{c} a_{H} \\ a_{L} \end{array}\right) = 
\left(\begin{array}{cc} k & \imath\,\theta' \\ -\imath\,\theta' & -k \end{array}\right) \;
\left(\begin{array}{c} a_{H} \\ a_{L} \end{array}\right). \label{eq:Hmatter}
\end{equation}
The prime denotes differentiation with respect to position $x$, the quantity $k$ is 
\begin{equation}
k(x) = k_V\;\frac{\sin\,2\theta_V}{\sin\,2\theta(x)}, \label{eq:k}
\end{equation}
with $k_V = \delta m^2 / (4 E)$ where $\delta m^2$ is the mass squared difference between the neutrino mass eigenstates, and 
\begin{equation}
\tan\,2\theta(x) = \frac{\sin\,2\theta_V}{\cos\,2\theta_V - V(x) / (2 k_V) }. \label{eq:tan2theta}
\end{equation}
Examination of equation (\ref{eq:k}) reveals that $k$ passes through minima whenever $\theta = \pi/4$ and these points in the profile are known as the `resonances'. The ratio $\gamma = k / |d\theta/dx|$ defining the adibaticity parameter is a measure of the strength of the mixing and the positions where $\gamma$ reaches minima are `points of maximal violation of adiabaticity'. It is here that the off-diagonal terms in equation (\ref{eq:Hmatter}) are most important and the mixing is at is strongest. Note, as pointed out by Friedland \cite{Friedland2001}, that in general the positions of `resonaces' and `points of maximal violation of adiabaticity' do not coincide and it is actually the latter that are more important for the evolution of the wavefunction. That said, throughout the remainder of the paper we will use these terms interchangeably. 
 
The Schrodinger equation forms a starting point from which the neutrino wavefunction emerging from the density profile can be determined. For 2-flavor mixing this is completley specified by calculating the `survival probability' i.e. the probability that a neutrino born as a particular flavor will emerge from the density profile as that same flavor. Unitarity provides the probability of detecting the counterpart flavor. Quite generally, if the rotation angle $\theta$ at the neutrino source is $\theta = \theta_{0}$ and the neutrino propagates to the vacuum then, after dropping the phase dependent terms, the flavor basis survival probability is \cite{K&P1989}
\begin{equation}
P(\nu_{\alpha} \rightarrow \nu_{\alpha}) = \frac{1}{2}\,\left[\; 1 + \cos 2\theta_{V}\,\cos 2\theta_{0}\,( 1 - 2 P_{C} ) \right]. \label{eq:PELZ}
\end{equation}
Here the quantity $P_{C}$ is known as the crossing probability. The crossing probability is a quantity defined in the matter basis and is the chance that an initial neutrino wavefunction transits from one matter eigenstate to the other. One obvious method to calculate $P_{C}$ is to simply integrate the Schrodinger equation in the matter basis. If $\gamma$ is always large as the neutrino propagates then the off-diagonal terms in equation (\ref{eq:Hmatter}) may be neglected, the integration of the Schrodinger equation is trivial and the wavefunction is said to evolve adibatically. There are also a handful of profiles where $P_{C}$ has an exact analytic solution \cite{K&P1989,K&T2001,Friedland2001}, the most well-known being the Landau-Zener result for the infinite linear profile: 
\begin{equation}
P_{C} = \exp[-\pi\gamma_{c}/2], \label{eq:PLZ}
\end{equation}
where $\gamma_{c}$ is the adibaticity parameter evaluated at the resonance. The Landau-Zener equation for $P_{C}$ possesses `troublesome pathologies' as discussed, and corrected, by Haxton \cite{Haxton1987}. 

But exact results are scant and often one finds that numeric integration of the Schrodinger equation for many interesting applications can be a frustrating exercise. As we mentioned previously, off-diagonal terms lead to oscillations and this is true even in the matter basis if the $\theta'$ term in equation (\ref{eq:Hmatter}) becomes large. Oscillatory solutions of differential equations obtained numerically are notorious for a gradual accumulation of error in both the phase and amplitude of the solution. A suitable change of variables can help to control some of these problems \cite{K&P1989} but even so, with a conventional solver, one usually has to be very aggressive with the error control in order to keep the solution accurate. This requirement can lead to very long run times.

In addition, the numeric integration of equation (\ref{eq:Hmatter}) is inefficient. With a convential differential equation solver the increments of the integration variable (here it is $x$) are necessarily smaller than the local oscillation length $(\sim 1/k)$. In regions of very high density, such as those found at the centers of supernova progenitor profiles, $k$ is very large and so the oscillation lengthscale will be very small. The differential equation solver will expend a great deal of time computing the wavefunction in such regions even though the large effective mass splitting indicates that the wavefunction is far from any resonance and $\gamma$ is large so that, in some regard, its evolution is both simple and uninteresting. 
Specialized methods, such as that by Petzold \cite{petzold81}, for highly oscillatory solutions of differential equations can help with the problem of small step sizes but their use may be limited by the requirement that any solution evolve adibatically.  

Due to these numeric problems, and motivated by a desire to comprehend the MSW effect, a number of alternate methods have been developed for calculating $P_{C}$. For example, one could estimate $P_{C}$ by using one of the exact results, most typically the Landau-Zener, if that approximation for the profile is adequate for the situation at hand. An alterantive approach would be to use the semi-analytic method by Balantekin \& Beacom \cite{BB1996} for arbitrary monotonic profiles. But for one reason or another these alternate approaches can break down or are difficult to automate. One such occurence is the case of multiple resonances and the computational methods used for the case of fluctuations in the solar profile \cite{LB1994,Loretietal1995,BFL1996,F&K2001} are much more sophisticated than brute-force application of a convential differential equation solver. 

In this paper we outline a new computational method for determining the neutrino wavefunction after its passage through a density profile. Our method undertakes the Monte Carlo integration of a scattering matrix and makes no assumptions with regard to adiabaticity or number of resonances and devotes the bulk of the computational time to the region around the point of maximal violation of adiabaticity. We derive our equations in section \S\ref{sec:equations} and discuss the Monte Carlo integration in section \S\ref{sec:calculating S}. We then consider the most important numerical difficulty in section \S\ref{sec:unitarity} before ending with applications of the technique to the density profile of the Sun and a density profile obtained from the evolution of supernova progenitor profile with a hydrodynamical calculation in section \S\ref{sec:applications}. Throughout this paper we will only consider two-flavor oscillations. In an appendix we expand on our ponderments of practical implementations.


\section{From The Schrodinger Equation To The Scattering Matrix} \label{sec:equations}

One persepctive on the evolution of the neutrino wavefunction through a density profile would be to regard the initial wavefunction as having been `scattered' so as to produce the emerging wavefunction. The scattering matrix that relates the outgoing wavefunction to the initial may be derived from the Schrodinger equation. As a first step in its determination we define a new variable $\phi$ via 
\begin{equation}
\frac{d\phi}{dx} = \frac{k}{\pi}, \label{eq:dphidx}
\end{equation}
so that
\begin{equation}
\phi = \frac{1}{\pi} \int k \, dx 
\end{equation}
and, secondly, we introduce a new basis, $\underline{b}$, related to the matter 
eigenstates via 
\begin{equation}
\left(\begin{array}{c} b_{H} \\ b_{L} \end{array}\right) 
= \left(\begin{array}{cc} e^{\imath\pi\phi} & 0 \\ 0 & e^{-\imath\pi\phi} \end{array}\right) \;
\left(\begin{array}{c} a_{H} \\ a_{L} \end{array}\right). \label{eq:beta definition}
\end{equation}
Equation (\ref{eq:dphidx}) allows us to change the independent variable from $x$ to $\phi$ and so measure distances in terms of this quantity. We see that $\phi$ has a physical interpretation as the number of half periods of the purely adiabatic solutions (i.e. $\theta^\prime = 0$) of the Schrodinger equation (\ref{eq:Hmatter}). Substitution of these definitions into equation (\ref{eq:Hmatter}) produces 
\begin{eqnarray}
\imath \frac{d}{d\phi} \left(\begin{array}{c} b_{H} \\ b_{L} \end{array}\right) & = &
\left(\begin{array}{cc} 0 & \imath\,\Gamma\,e^{2\imath\pi\phi} \\ -\imath\,\Gamma\,e^{-2\imath\pi\phi} & 0 \end{array}\right)\;\left(\begin{array}{c} b_{H} \\ b_{L} \end{array}\right) \nonumber \\
& = & H(\phi)\;\left(\begin{array}{c} b_{H} \\ b_{L} \end{array}\right) 
\label{eq:Hbeta}
\end{eqnarray}
where $\Gamma$ is the related to the adiabaticity parameter, 
\begin{equation} 
\Gamma = \pi/\gamma. \label{eq:Gamma}
\end{equation}
Note that this definition indicates that a wavefunction will evolve non-adiabaticly if $\Gamma >>1$. The change in basis allows us to focus the problem on the non-adiabatic part of the solution by which we mean that portion that jumps from one matter eigenstate to the other because, in this new basis, the Schrodinger equation is now purely off-diagonal.
By integrating equation (\ref{eq:Hbeta}) we obtain 
\begin{equation} 
\left(\begin{array}{c} b_{H} \\ b_{L} \end{array}\right)
=\left(\begin{array}{c} b_{H0} \\ b_{L0} \end{array}\right)
-\imath\;\int_{0}^{\Phi} \; d\phi_{1} H(\phi_{1}) \;
\left(\begin{array}{c} b_{H} \\ b_{L} \end{array}\right),
\end{equation}
and repeated substitution of this result into itself yields 
\begin{widetext}
\begin{eqnarray} 
\left(\begin{array}{c} b_{H} \\ b_{L} \end{array}\right)
& = & \left(\begin{array}{c} b_{H0} \\ b_{L0} \end{array}\right)
-\imath\;\int_{0}^{\Phi} \; d\phi_{1} \;H_{1} \,
\left(\begin{array}{c} b_{H0} \\ b_{L0} \end{array}\right) +  (-\imath)^{2}\;
\int_{0}^{\Phi} \; d\phi_{1} \; H_{1} \, \int_{0}^{\phi_{1}} \; d\phi_{2} \;H_{2} \,
\left(\begin{array}{c} b_{H0} \\ b_{L0} \end{array}\right) +  \ldots \\
& = & \left\{ 1 -\imath\;\int_{0}^{\Phi} \; d\phi_{1} \;H_{1} \, 
+  (-\imath)^{2}\; \int_{0}^{\Phi} \; d\phi_{1} \; H_{1} \, \int_{0}^{\phi_{1}} \; d\phi_{2} \;H_{2} \,
+  \ldots \right\} \; \left(\begin{array}{c} b_{H0} \\ b_{L0} \end{array}\right) \label{eq:integral}
\end{eqnarray}
\end{widetext}
where the subscripts on the $H$'s indicate their argument with respect to $\phi$ by which we mean $H_{i} = H(\phi_{i})$. This equation defines the scattering matrix $S(\Phi)$ in this basis as 
\begin{equation} 
\left(\begin{array}{c} b_{H} \\ b_{L} \end{array}\right) 
 =  S(\Phi) \left(\begin{array}{c} b_{H0} \\ b_{L0} \end{array}\right).
\end{equation}
The upper limits on the integrals appearing in equation (\ref{eq:integral}) indicate the space/time ordering ($\phi$ is a monotonically increasing function of $x$) but we may change all the upper limits to $\Phi$ by using such identities as 
\begin{eqnarray}
& & \int_{0}^{\Phi} d\phi_{1} \; H_{1} \int_{0}^{\phi_{1}} \; d\phi_{2} \;H_{2} 
= \frac{1}{2!} \,\int_{0}^{\Phi}\;d\phi_{1}\,\int_{0}^{\Phi}\;d\phi_{2}\, \nonumber \\
& & \;\;\;\;\;\;\;\; \left\{ H_{1}\,H_{2} \Theta(\phi_{1}-\phi_{2}) + H_{2}\,H_{1} \Theta(\phi_{2}-\phi_{1}) \right\}. \label{eq:S2timeorder}
\end{eqnarray}
where $\Theta(\phi_{1}-\phi_{2})$ is the step function. This, and similar identities for the the higher order multiple integrals, allows us to write $S(\Phi)$ as 
\begin{widetext}
\begin{equation}
S(\Phi) = 1 + (-\imath)\int_{0}^{\Phi}d\phi_{1}\,H_{1} + \frac{(-\imath)^{2}}{2!}\int_{0}^{\Phi}d\phi_{1} \int_{0}^{\Phi}d\phi_{2}\,\mathbb{T}(H_{1}\,H_{2}) + \frac{(-\imath)^{3}}{3!}\int_{0}^{\Phi}d\phi_{1} \int_{0}^{\Phi}d\phi_{2}\int_{0}^{\Phi}d\phi_{3}\,\mathbb{T}(H_{1}\,H_{2}\,H_{3}) + \ldots, \label{eq:Sproduct}
\end{equation}
\end{widetext}
where $\mathbb{T}$ is the $\phi$-ordered product. With the scattering matrix defined we describe our
approach to its calculation. 

 
\section{Monte Carlo Calculations For The Scattering Matrix} \label{sec:calculating S}

The conversion from a differential to an integral equation means that completely different numerical algorithms must be applied. The number of terms that one may have to include in equation (\ref{eq:Sproduct}) to achieve sufficient accuracy, and the fact that $H(\phi)$ involves an $e^{2\imath\pi\phi}$ oscillatory terms, likely precludes any approach other than a Monte Carlo integration. Though Monte Carlo methods are often regarded as a last resort their usefulness becomes apparent when either the boundaries of the integration region are very complicated or, as in this case, when the dimensionality of the integration measure means that more sophisticated algorithms will not produce a result in a respectable amount of time.

The quantities we select randomly are $\phi_{i}$ to be drawn from a probability distribution $P(\phi)$. Na\"{i}vely we could pick values for $\phi_{1},\;\phi_{2},...$ from a uniform range between $0$ and $\Phi$ but the structure of equation (\ref{eq:Sproduct}) shows that this would be inefficient because the Hamiltonians are all proportional to $\Gamma$ and this quantity is largest in the region close to the resonance. 
This would suggest that we should select $P(\phi) \propto \Gamma$ and hence use importance sampling for the $\phi$'s. This would be fine for the case of only one resonance but if there are multiple resonances we encounter problems due to the fact that $\Gamma \propto \theta'$. If $\theta'$ ever switches sign then $P(\phi)$ would switch sign and, over some portion of the profile, we would have a negative probability distribution. So instead we use $P(\phi) \propto |\Gamma|$.

\begin{figure}[h]
\begin{center}
\includegraphics[width=8.6cm]{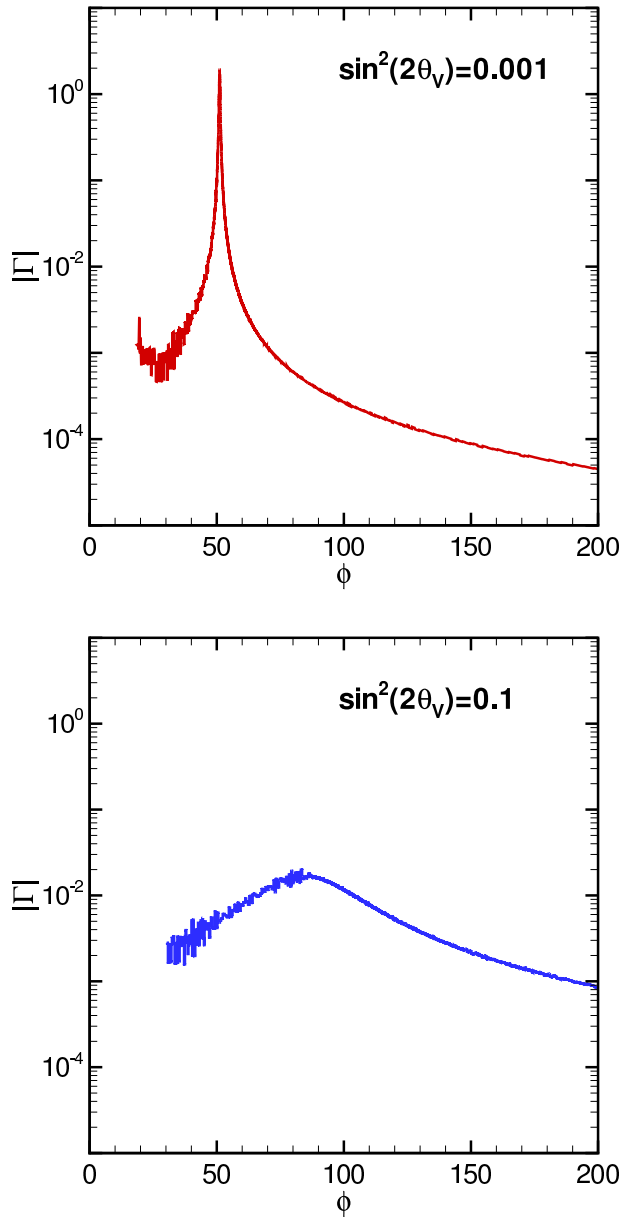} 
\caption{The absolute value of the inverse adiabaticity parameter $\Gamma$ as a function of the period counting variable $\phi$. The density profile is the BS2005-AGS,OP Standard Solar Model \cite{Bahcall2005} and we selected $\delta m^{2} = 3 \times 10^{-5} \;{\rm eV^{2}}$, $E=10 \;{\rm MeV}$. In the top panel we used $\sin^{2}(2\,\theta_{V})=0.001$ and for the bottom $\sin^{2}(2\,\theta_{V})=0.1$. \label{fig:Gammasolar} }
\end{center}
\end{figure}
To illustrate just how sharply peaked $|\Gamma|(\phi)$ can be
we show in figure (\ref{fig:Gammasolar}) this function for the BS2005-AGS,OP Standard Solar Model \cite{Bahcall2005} for two different values of $\sin^2 2 \theta_V$. For the upper panel we selected $\delta m^{2} = 3 \times 10^{-5} \;{\rm eV^{2}}$, $E=10 \;{\rm MeV}$ and $\sin^{2}(2\,\theta_{V})=0.001$ which, as the figure indicates since the peak is $|\Gamma| > 1$, means that the resonance is non-adiabatic. The point of maximal violation of adiabaticity is where $\gamma \propto 1/\Gamma$ reaches it's minimum value so by using $P(\phi) \propto |\Gamma|$ as the probability distribution for $\phi$ we concentrate our efforts around this point. The bottom panel shows $|\Gamma|$ for the case of $\sin^2 2 \theta_ V = 0.1$.  For this larger value, $|\Gamma|$ is less sharply peaked, the wavefunction evolves adibatically and the values of $\phi$ we obtain from this probability distribution are spread over a broad range. 

Before we proceed the probability distribution must be normalized. The normalization $A$ for the distribution, $P(\phi) = A\,|\Gamma|$, is simply
\begin{equation}
1/A = \int_{0}^{\Phi}\;d\phi'\,|\Gamma| = \int_{0}^{x}\;dx' |\frac{d\theta}{dx'}|. \label{eq:normalization}
\end{equation}
With $P(\phi)$ identified we can pull out from $H(\phi)$ the probability distribution $P(\phi)$ and define a reduced Hamiltonian $h(\phi)$ as $H(\phi) = P(\phi)\,h(\phi)$; written explicitly $h(\phi)$ is 
\begin{equation}
h(\phi) = \frac{sign[\Gamma(\phi)]}{A}\;\left(\begin{array}{cc} 0 & \imath\,\,e^{2\imath\pi\phi} \\ -\imath\,e^{-2\imath\pi\phi} & 0 \end{array}\right). \label{eq:hreduced}
\end{equation}

The definition for the scattering matrix in equation (\ref{eq:Sproduct}) is a sum of multiple integrals but by utilizing 
the identity 
\begin{equation}
1 = \int_{0}^{\Phi}\;P(\phi')\,d\phi' 
\end{equation}
the sum can be collapsed down to a single multiple integral albeit one with infinite dimensionality for its measure:
\begin{widetext}
\begin{equation}
S = \left( \prod_{i=1}^{\infty} \int_{0}^{\Phi}\,P(\phi_{i})\,d\phi_{i} \right) 
\left\{ 1 + (-\imath)\,h_{1} + \frac{(-\imath)^{2}}{2!}\,\mathbb{T}(h_{1}\,h_{2}) + \frac{(-\imath)^{3}}{3!}\,\mathbb{T}(h_{1}\,h_{2}\,h_{3}) + \ldots \right\}. \label{eq:Sreduced}
\end{equation}
\end{widetext}
This expression for the scattering matrix is more useful from a practical standpoint because it allows us to reuse values of $\phi$. The scattering matrix is therefore the expectation value of the quantity $s$ where 
\begin{eqnarray}
s & = & 1 + (-\imath)\,h_{1} + \frac{(-\imath)^{2}}{2!}\,\mathbb{T}(h_{1}\,h_{2}) \nonumber \\
  &   & \;\; + \frac{(-\imath)^{3}}{3!}\,\mathbb{T}(h_{1}\,h_{2}\,h_{3}) + \ldots \label{eq:s} 
\end{eqnarray}
Our algorithm for the calculation of $S$ is to simply form a set with $N_{T}$ samples of the matrix $s$ and average them. 


\section{The Crossing Probability From The Scattering Matrix}

The scattering matrix possesses a simple structure defined by two complex numbers $\alpha$ and $\beta$ so that
\begin{equation}
S=\left(\begin{array}{cc} \alpha & \beta \\ -\beta^{\ast} & \alpha^{\ast} \end{array}\right).
\end{equation}
It is tempting to regard $\alpha$ and $\beta$ as Cayley-Klein parameters but, as we shall discuss below, 
the Monte Carlo does not guarantee that $S^{\dagger}\,S =1$ or, equivalently, $|\alpha|^{2} + |\beta|^{2} =1$

Once $S$ is calculated  we apply $S$ to the initial wavefunction $\underline{b}(0)$ in order to determine $\underline{b}(\Phi)$, i.e $\underline{b}(\Phi) = S(\Phi)\,\underline{b}(0)$. 
The crossing probability, $P_{C}$, is the chance that a wavefunction prepared in a pure matter eigenstate 
has transited to the other matter eigenstate as it emerges from the profile. Our scattering matrix is defined in the 
$\underline{b}$ basis, not the matter, $\underline{a}$, basis and these are related by the expression in equation (\ref{eq:beta definition}). The crossing probability is thus
\begin{eqnarray}
P_{C}^{\beta} & = & \left(\begin{array}{cc} 1 & 0 \end{array}\right) S^{\dagger}  
\left(\begin{array}{cc} e^{\imath\pi\Phi} & 0 \\ 0 & e^{-\imath\pi\Phi} \end{array}\right)
\left(\begin{array}{c} 0 \\ 1 \end{array}\right) \nonumber \\
& & \;\;\;\;\; \left(\begin{array}{cc} 0 & 1 \end{array}\right) 
\left(\begin{array}{cc} e^{-\imath\pi\Phi} & 0 \\ 0 & e^{\imath\pi\Phi} \end{array}\right)
S \left(\begin{array}{c} 1 \\ 0 \end{array}\right) \\
& = & |\beta|^{2} \label{eq:PCb}.
\end{eqnarray}
The superscript upon $P^{\beta}_{C}$ is to remind the reader of the second line of this equation. 

We may also define $P_{C}$ as being the difference from unity of the probability that a wavefunction prepared in a pure matter eigenstate survives as that same matter eigenstate as it emerges from the profile: i.e.
\begin{widetext}
\begin{eqnarray}
P_{C}^{\alpha} & = & 1- \left(\begin{array}{cc} 1 & 0 \end{array}\right) S^{\dagger}  
\left(\begin{array}{cc} e^{\imath\pi\Phi} & 0 \\ 0 & e^{-\imath\pi\Phi} \end{array}\right)
\left(\begin{array}{c} 1 \\ 0 \end{array}\right)
\left(\begin{array}{cc} 1 & 0 \end{array}\right) 
\left(\begin{array}{cc} e^{-\imath\pi\Phi} & 0 \\ 0 & e^{\imath\pi\Phi} \end{array}\right)
S \left(\begin{array}{c} 1 \\ 0 \end{array}\right) \\
& = & 1 - |\alpha|^{2} \label{eq:PCa}
\end{eqnarray}
\end{widetext}
Again, the superscript upon $P^{\alpha}_{C}$ is to remind the reader of the second line of this equation. 

Thus our scattering matrix can be used to construct two values for $P_{C}$ and if $S$ were unitary then
they would be equal. The difference between them is due to finite sampling and is
the subject of the next section.


\section{The distributions of $P_{C}^{\alpha}$ and $P_{C}^{\beta}$ and the unitarity of $S$} \label{sec:unitarity}

After execution of the Monte Carlo algorithm for a given profile and mixing parameters, 
one obtains a scattering matrix $S$ from which $P_{C}^{\alpha}$ and $P_{C}^{\beta}$ 
can be formed. The scattering matrix calculated by this method does not, in general, 
guarantee that the identity $S^{\dagger}\,S -1 =0$ is satisfied. 
This is equivalent to the statements that $|\alpha|^{2}+|\beta|^{2} -1 \neq 0$ 
and $P_{C}^{\beta} - P_{C}^{\alpha} \neq 0$. Thus $\alpha$ and $\beta$ are 
not Cayley-Klein parameters and the two calculated crossing probabilities are not exactly equal. 
Also, if the calculation for a given profile and mixing paramaters is repeated then we obtain a new scattering 
matrix and new values for $P_{C}^{\alpha}$ and $P_{C}^{\beta}$. 
The difference between the two crossing probabilities for a given run and their  
change from one run to the next arises because we only construct a finite set of 
samples of $s$. Only in the limit of an infinite number of samples would 
$P_{C}^{\alpha}$ and $P_{C}^{\beta}$ be exactly equal and our calcualtion give 
the same result every time. 

We stress that this behavior is not a fundamental flaw of the Monte Carlo 
technique but rather a numeric issue related to the usual lack of infinite computing resources. 
For this reason one must be content with values for $P_{C}^{\alpha}$ 
and $P_{C}^{\beta}$ that differ from the true crossing probability and from each other.  
With the cautionary note that what follows is specific to our implementation of the algorithm 
and the test problems we selected, we try and provide some guidance on how to obtain
the most accurate calculation in the least computational time.

\begin{figure}[htbp]
\begin{center}
\includegraphics[width=8.6cm]{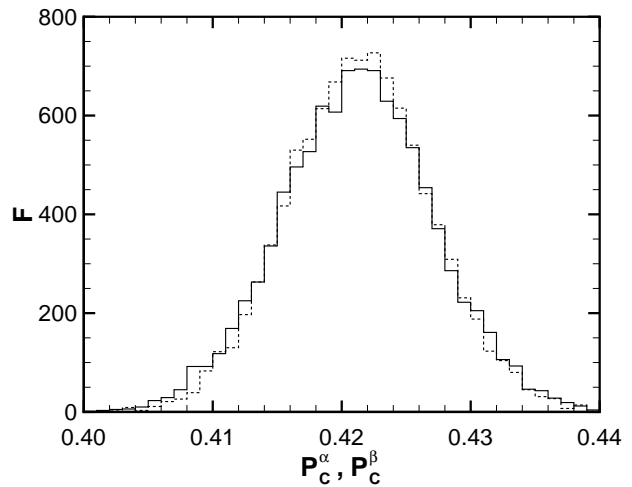} \caption{The frequency distribution
of $P_{C}^{\alpha}$ (solid) and $P_{C}^{\beta}$ (dashed) of 10,000 results 
from the Monte Carlo calculation using the BS2005-AGS,OP Standard Solar Model 
\cite{Bahcall2005}, $\delta m^{2} = 3 \times 10^{-5} \;{\rm eV^{2}}$, $E=10 
\;{\rm MeV}$ and $\sin^{2}(2\,\theta_{V})=0.001$ as the physical 
parameters. The number of trials is $N_{T}=10^{4}$. \label{fig:F091} }
\end{center}
\end{figure}
The values of $P_{C}^{\alpha}$ and $P_{C}^{\beta}$ obtained from a given calculation 
are drawn from parent distributions that, in general, are unique to the particular 
profile, mixing parameters and also the implementation of the algorithm. 
These parent distributions may be reconstructed by repeating the calculation for 
$P_{C}^{\alpha}$ and $P_{C}^{\beta}$ until 
a sufficiently large sample of results has been extracted. 
As an example, the frequency distributions of $P_{C}^{\alpha}$ and $P_{C}^{\beta}$ 
for the case of $10\;{\rm MeV}$ neutrino passing through the
BS2005-AGS,OP Standard Solar Model density profile with 
$\delta m^{2} = 3 \times 10^{-5} \;{\rm eV^{2}}$ and $\sin^{2}(2\,\theta_{V})=0.001$
are shown in figure (\ref{fig:F091}). 
The initial impression is that the distributions for $P_{C}^{\alpha}$ and $P_{C}^{\beta}$ are 
both Gaussian with similar means and variances and indeed the $d$-statistic from a
Kolmogorov-Smirnov test (using the Lilliefors \cite{Lillie67} critical values) verifies this 
conclusion. 
\begin{figure}[htbp]
\begin{center}
\includegraphics[width=8.6cm]{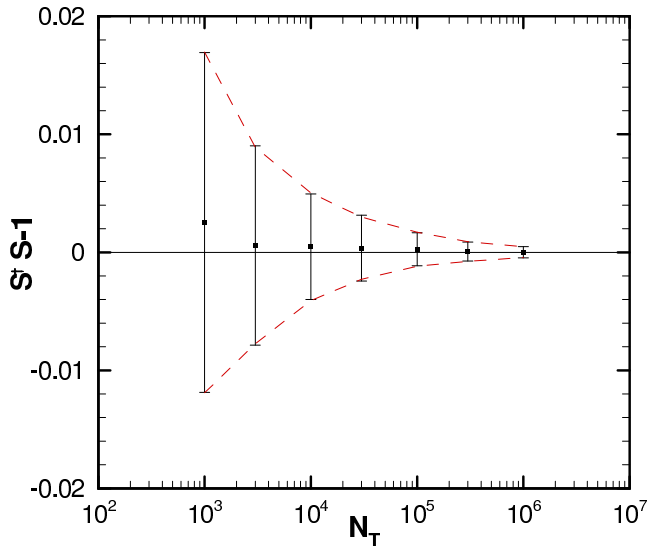} \caption{The mean value of 
$\overline{S^{\dagger}\,S-1}$ indicated by the squares as a 
function of $N_{T}$ for the calculation outlined in the text. 
The error bars are not the error in the mean but rather indicate 
the $1-\sigma$ spread in the values of $S^{\dagger}\,S-1$. The 
dashed lines are $\overline{S^{\dagger}\,S-1} \pm 
C/\sqrt{N_{T}}$ with $C=0.468$. \label{fig:SdagS} }
\end{center}
\end{figure}
The departure from unitarity is affected by the number of trials 
that go into the calculation of $S$. 
In figure (\ref{fig:SdagS}) we plot the mean value of 
$\overline{S^{\dagger}\,S-1}$ and the $1-\sigma$ spread of the 
sample for various values of $N_{T}$. Again, the calculation is for a 
neutrino passing through the BS2005-AGS,OP Standard Solar Model 
\cite{Bahcall2005} density profile and for 
$\delta m^{2} = 3 \times 10^{-5} \;{\rm eV^{2}}$, $E=10 
\;{\rm MeV}$ and $\sin^{2}(2\,\theta_{V})=0.001$. 
The mean values of 
$\overline{S^{\dagger}\,S-1}$ all lie above zero indicating that the mean
value of $P_{C}^{\beta}$ is apparently slightly larger than the mean for $P_{C}^{\alpha}$ 
but that this difference disappears as $N_{T}$ increases. The error bars 
on each point are the $1-\sigma$ spread in $S^{\dagger}\,S-1$ and these 
clearly diminish as $N_{T}$ increases. We find that the size of the 
error bars follows a trend proportional to $1/\sqrt{N_{T}}$ as indicated 
by the dashed lines in the figure. The figure indicates 
unitarity is achieved in the limit when the number of 
samples of the scattering matrix becomes infinite.  

\begin{figure}[htbp]
\begin{center}
\includegraphics[width=8.6cm]{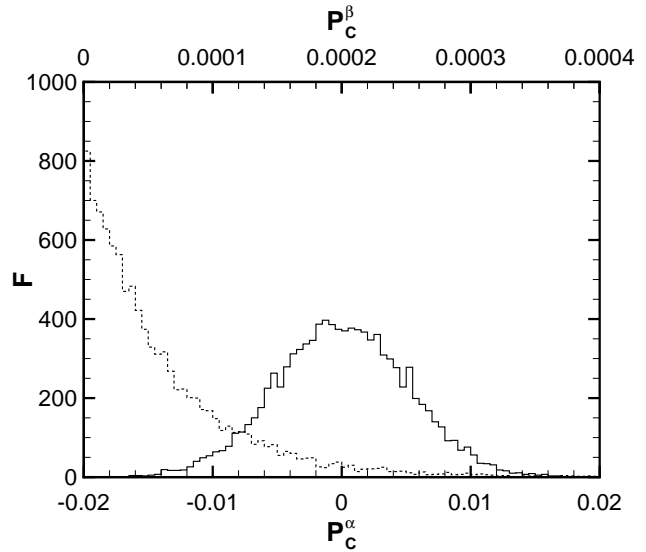} \caption{The frequency distribution
of $P_{C}^{\alpha}$ (solid) and $P_{C}^{\beta}$ (dashed) of 10,000 results 
from the Monte Carlo calculation using the BS2005-AGS,OP Standard Solar Model 
\cite{Bahcall2005}, $\delta m^{2} = 3 \times 10^{-5} \;{\rm eV^{2}}$, $E=10 
\;{\rm MeV}$ and $\sin^{2}(2\,\theta_{V})=0.1$ as the physical 
parameters. The number of trials is again $N_{T}=10^{4}$. \label{fig:F922} }
\end{center}
\end{figure}
But Gaussianity in the distributions for $P_{C}^{\alpha}$ and $P_{C}^{\beta}$ 
does not always occur and should not be taken for granted. 
If we consider a case where $P_{C}$ is close to zero non-Gaussianity becomes apparent. 
In figure (\ref{fig:F922}) we show the frequency distributions for the 
case of $\delta m^{2} = 3 \times 10^{-5} \;{\rm eV^{2}}$, 
$E=10 \;{\rm MeV}$ and $\sin^{2}(2\,\theta_{V})=0.1$. 
The distribution for $P_{C}^{\alpha}$ remains Gaussian but we immediately notice 
that the distribution for $P_{C}^{\beta}$ has changed and we find that 
a Gamma distribution with an $\alpha$ parameter that is close to unity 
is a good fit. The figure also shows that \emph{negative} values $P_{C}^{\alpha}$ can be obtained whereas $P_{C}^{\beta}$ is always positive. This result arises due to the definitions in equation (\ref{eq:PCb}) and (\ref{eq:PCa}): values of $P_{C}^{\beta}$ less than zero are not allowed but there is no similar restriction for $P_{C}^{\alpha}$. 
Note also the very different widths of the distributions: 
the width of the $P_{C}^{\alpha}$ distribution is similar to that in figure (\ref{fig:F091})
but the width of $P_{C}^{\beta}$ has shrunk considerably. This difference in the widths of the two crossing
probabilities would indicate that the deviation from unitarity, $S^{\dagger}\,S-1$, will be dominated by the spread in 
$P_{C}^{\alpha}$, the values of $P_{C}^{\beta}$ having such a small variance. Thus, when $P_{C}$ is close to zero $P_{C}^{\beta}$ is much more accurately calculated than $P_{C}^{\alpha}$. 
We also find that for this test case, the width of the two distributions varies with $N_{T}$ in different fashions.
For $P_{C}^{\alpha}$ the spread again varies as $1/\sqrt{N_{T}}$ but the width of the $P_{C}^{\beta}$ distribution 
now behaves as $1/N_{T}$. From additional test cases we found that that when $P_{C}$ approaches unity 
it is $P_{C}^{\alpha}$ that is the more accurately calcualted. Our experience has also shown that in some circumstances the distributions for $P_{C}^{\alpha}$ and $P_{C}^{\beta}$ 
can also change shape as $N_{T}$ is varied: for small $N_{T}$ the distibution may be like a 
Gamma distribution with a modest $\alpha$ parameter but then will morph to something closer to a Gaussian distribution as $N_{T}$ increases. 

These results hint at the interesting underlying numerics of this Monte Carlo approach but they also introduce some 
confusion into what would be a reasonable modus operandi. The parent distributions for $P_{C}^{\alpha}$ and $P_{C}^{\beta}$ are not, in general, the same and we do not know a priori their shape or if they are similar. This would seem to preclude combining the results for $P_{C}^{\alpha}$ and $P_{C}^{\beta}$ in some way so as to obtain 
a more accurate result. The accuracy of the results depend upon $N_{T}$ but in a way that varies as we change the profile and mixing parameters. Before we do the calculation we do not know how large we must make $N_{T}$ to reach our intended level of accuracy. In practice we adopted a `worst case scneario' approach whereby we calculate both 
$P_{C}^{\alpha}$ and $P_{C}^{\beta}$ assuming that the accuracy varies as $1/\sqrt{N_{T}}$. One would then require $N_{T} \sim 10^{6}$ trials to reach a level of accuracy of $\sim 0.1\%$. We then used 
$P_{C}^{\beta}$ for the crossing probability if $P_{C} \leq 0.5$ and $P_{C}^{\alpha}$ otherwise. 
As we said, the shape of the distributions for $P_{C}^{\alpha}$ and $P_{C}^{\beta}$ can vary with the number of samples so breaking up the $N_{T}$ trials into a number of smaller runs (e.g. $10$ runs with $10^{5}$ samples in each), 
calculating $P_{C}^{\alpha}$ and $P_{C}^{\beta}$ from each run and then averaging the results must also be approached with caution. To avoid potential bias in such a procedure we only accepted the result from the one run with the full number of trials we specified. Though this conservative approach has the drawback that the runtime of our code may be longer than necessary the results always achieve our desired level of accuracy and often considerably so. 


\section{Example Calculations} \label{sec:applications}

We finish with three applications of our method. We first demonstrate the method with
a calculation of the survival probability of electron neutrinos using the solar density profile
and two different values of $\sin^2 2 \theta_V$. We then go on to use profile from
an aspherical supernova simulation, which involves multiple resonances.

\begin{figure}[htbp]
\begin{center}
\includegraphics[width=8.6cm]{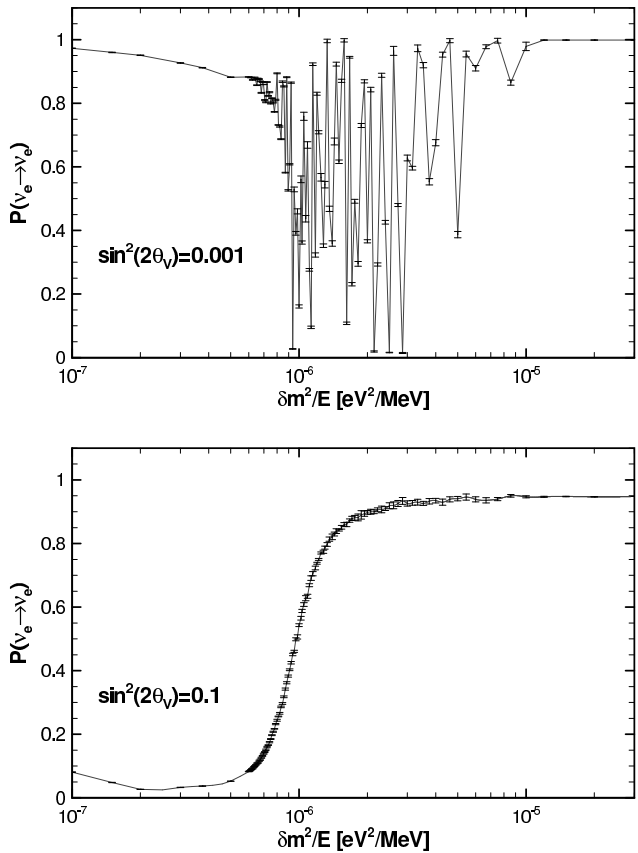} 
\caption{The neutrino survival probability through the Standard Solar Model density profile as a function of $\delta m^{2} /E$. In the top panel $\sin^{2}(2\,\theta_{V})=0.001$, in the bottom $\sin^{2}(2\,\theta_{V})=0.1$. The source of neutrinos is located at $3/10\;R_{\odot}$ and the neutrinos propagate back through the core and out the other side. The error bars on each point are the rms spread in the results from 8 repetitions. The Gaussian estimator leads to a bias so the accuracy should be regarded as illustrative. \label{fig:Psolar} }
\end{center}
\end{figure}
The passage of neutrinos through the solar density profile is a well studied problem 
and therefore there are a number of already published calculations.
In figure (\ref{fig:Psolar}) we calculate the survival probability of electron neutrinos over a spectrum in energy through the BS2005-AGS,OP Standard Solar Model \cite{Bahcall2005}. For these figures we select either $\sin^{2}(2\,\theta_{V})=0.001$ or $\sin^{2}(2\,\theta_{V})=0.1$. The source of neutrinos is located at $3/10$ of the solar radius and they propagate back through the core and emerge the other side. These figures agree those of 
Haxton \cite{Haxton1987} for the same calculation. In these calculations the lower energy neutrinos experience a double resonance while the higher energy neutrinos experience only one. This changeover is seen in the bottom panel of figure (\ref{fig:Psolar}) where the survival probability transits from $\sim 0$ to $1$ at $\delta m^{2} /E \sim 10^{-6}\;{\rm eV^{2}/MeV}$. The top panel in figure (\ref{fig:Psolar}) exhibits rapid fluctuations in the survival probability (which are by no means resolved with energy spacing we used) and indicate phase effects as discussed in \cite{K&P1989}. These features in the figure demonstrate that the Monte Carlo is capable of reproducing the results of other calculations. 

The measured value of $\theta_{V,solar}$ is larger than what we
have used in our example calculations here, 
and therefore neutrinos from the sun go through adiabatic neutrino
flavor transformation.  However, the value of $\theta_{13}$ is yet unknown.  This angle
will determine the degree of flavor transformation in the core collapse supernova.
\begin{figure}[htbp]
\begin{center}
\includegraphics[width=8.6cm]{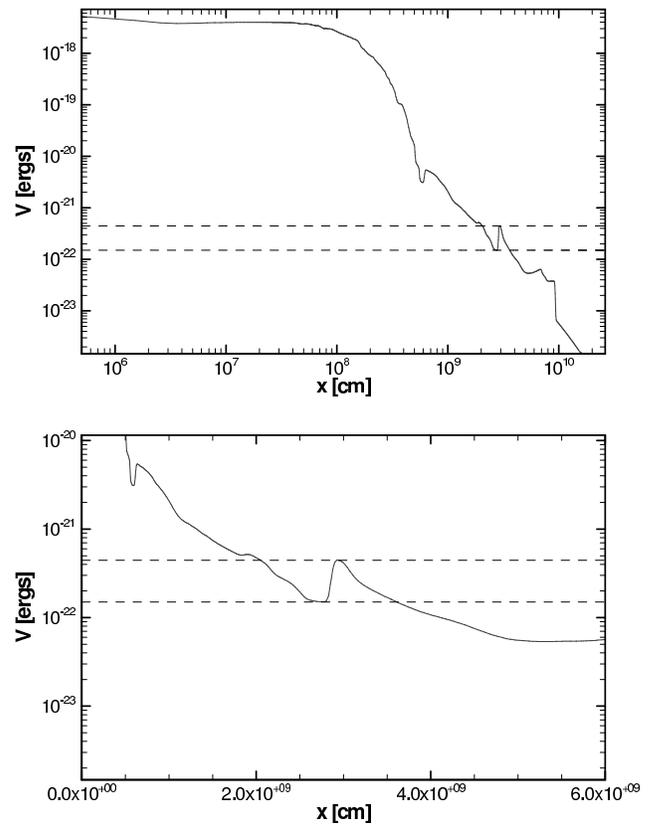} 
\caption{The electron neutrino potential energy, $V=\sqrt{2}\,G_{F} n_{e}$, as a function of radial distance for the model discussed in the text. The upper figure is the entire profile, the lower focuses upon that portion up to $6\times 10^{9}\;{\rm cm}$. In both panels the dashed lines indicate the resonance potential energies for $5.4\;{\rm MeV}$ (upper) and $16\;{\rm MeV}$ (lower) neutrinos indicating that neutrinos with energies between these values will experience a triple resonance. The mass splitting is chosen to be $\delta m^{2} = 3\times 10^{-3}\;{\rm eV^{2}}$ and $\sin^{2}(2\theta_{V}) = 4\times 10^{-4}$. \label{fig:3resV} }
\end{center}
\end{figure}
\begin{figure}[htbp]
\begin{center}
\includegraphics[width=8.6cm]{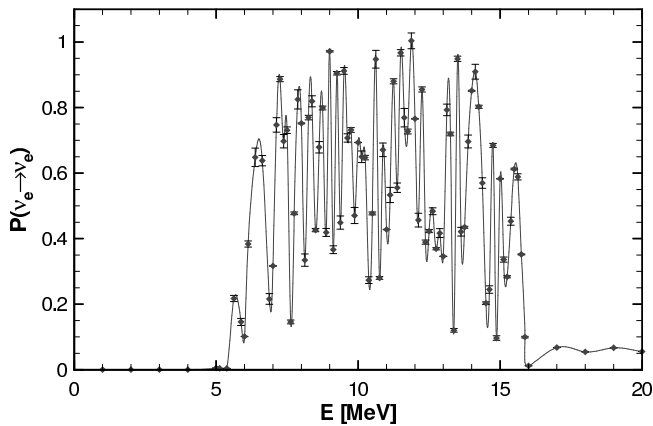} 
\caption{The neutrino survival probability through the density profile shown in figure (\ref{fig:3resV}) as a function of the neutrino energy. For this calculation $\delta m^{2} = 3\times 10^{-3}\;{\rm eV^{2}}$ and $\sin^{2}(2\theta_{V}) = 4\times 10^{-4}$. Again, the error bars on each point are the rms spread in the results from 8 repetitions of the calculation using the Gaussian estimator. Though this estimator is biased they indicate the accuracry of the result. \label{fig:P3res} }
\end{center}
\end{figure}
Therefore,  we consider finally  the more complicated profile shown in figure (\ref{fig:3resV}). This profile is a product of the evolution of a supernova progenitor model using the VHI  hydrodynamical code. An $\ell =2$ spherical harmonic velocity perturbation was inserted by hand into the u13.2 progenitor profile of Heger \cite{Heger} to cause the star to explode asymmetrically. As a consequence of the asphericity several density minima were produced and the profile shown is a radial slice through the model $9\;{\rm s}$ after the bounce. 
We select $\delta m^{2} = 3\times 10^{-3}\;{\rm eV^{2}}$ and $\sin^{2}(2\theta_{V}) = 4\times 10^{-4}$ and find that neutrinos with energies between $5.4\;{\rm MeV}$ and $16\;{\rm MeV}$ will experience a triple resonance, this region is magnified in the lower panel of figure (\ref{fig:3resV}). The results of the calculation are shown in figure (\ref{fig:P3res}). Between $5.4\;{\rm MeV}$ and $16\;{\rm MeV}$ the survival probability, again, exhibits phase effects. At present
we only wish to illustrate a potential  use of the technique presented here for the
case of multiple resonances.
The calculation leading to this profile will be discussed elsewhere along with a more detailed studied of the observable consequences of multiple resonances from aspherical supernova
explosions \cite{BBKM2005}. 


\section{Summary} \label{sec:summary}

We have shown that the effects of matter upon the propagation of neutrinos may be described as the scattering of an initial neutrino wavefunction permitting us to exchange the differential Schrodinger equation for an integral equation for the scattering matrix. In this formulism we are able to avoid the numerical difficulties associated with oscillatory solutions to differential equations by, instead, using Monte Carlo integrators that focus the calculation onto the most important aspects of the problem. Though slow to converge compared to more sophisticated methods, and possessing inherent numerical error due to finite sampling, Monte Carlo integrators have the advantage of easily controlled runtimes and the numerical errors are both understood and ameliorated by repetition. This technique may be useful in a number of interesting density profiles that are difficult to work with using traditional techniques, 
particularily those involving multiple resonance regions, such as in a core collapse supernovae.


\vspace{1.cm} This work was supported at NCSU by the U.S. Department of Energy under grant DE-FG02-02ER41216 
and at UMn under grant DE-FG02-87ER40328. 


\appendix

\section{Practical Considerations} \label{sec:practical}

As we described in section \S\ref{sec:calculating S}, the scattering matrix is 
found to be the average of a set of samples for $s$. As a reminder, $s$ is given by
\begin{eqnarray}
s & = & 1 + (-\imath)\,h_{1} + \frac{(-\imath)^{2}}{2!}\,\mathbb{T}(h_{1}\,h_{2}) \nonumber \\
& & \;\; + \frac{(-\imath)^{3}}{3!}\,\mathbb{T}(h_{1}\,h_{2}\,h_{3}) + \ldots \label{eq:sappendix} \\
& = & \sum_{i=0}^{\infty} s_{i} 
\end{eqnarray}
and the subscripts on the reduced Hamiltonians $h$ indicate the $\phi$ argument 
by which we mean $h_{i} = h(\phi_{i})$ and $h(\phi)$ is given by equation (\ref{eq:hreduced}).
Constructing a set of $s$ to average is the principle task of the algorithm.
It is not our intention to proscribe a recipe for the construction 
of $s$, and the reader can find many additional runtime savings that are not discussed here, 
but rather we outline some considerations that may be useful. 

\subsection{Truncating the series}

Formally $s$ is the sum of an infinite number of terms but in 
practice we must truncate the series at some order $N_{S}$. The 
basis for selection of $N_{S}$ comes from noticing that the terms 
in $s$ are proportional to a unitary matrix and a weighting 
factor $w$ with 
\begin{equation}
w_{i} = \frac{1}{i!A^{i}}.
\end{equation} 
We can set a value for $N_{S}$ by requiring that the 
weight of the terms we retain are larger than some specified level 
$\epsilon_{S}$; that is, $w_{N_{S}} \geq \epsilon_{S}$. 
Since the weights are inversely related to the normalization the 
smaller the value of $A$ then then larger $N_{S}$. Small value of 
$A$, as seen in equation (\ref{eq:normalization}), occur for 
greater differences between the initial and final rotation angles 
across a resonance and/or the greater the number of zeros for 
$\Gamma$. The value of $\epsilon_{S}$ should be sufficiently small that 
the numerical error in the values of $P_{C}^{\alpha}$ and $P_{C}^{\beta}$ 
should be dominated by the finite sampling error otherwise the crossing
probabilities would contain a systematic error due to this truncation. 
For a desired level of accuracy of $\sim 0.1\%$ in $P_{C}^{\alpha}$ and $P_{C}^{\beta}$ 
we found that $\epsilon_{S}\sim 10^{-4}$ was sufficient.

\subsection{Generating random values of $\phi$}

With $N_{S}$ chosen the structure of equation (\ref{eq:sappendix}) 
indicates we need $N_{S}$ values of $\phi$ to compute $s$. 
The most efficient method for obtaining a sequence 
of $\phi$'s from the probability distribution $P(\phi)$ is to 
relate $P(\phi)$ to the uniform distribution so that one may use 
a pseudo-random number generator. This is achieved by calculating the 
accumulated probability, $F(\phi)$, from 
\begin{equation}
F(\phi) = \int_{0}^{\phi} A|\Gamma(\phi')| \,d\phi' 
\end{equation}
and then inversion of the relationship to form $\phi(F)$. The requirement that 
$F(\Phi)=1$ sets the normalization $A$ as shown in equation (\ref{eq:normalization}). 
After substituting the definition of $\Gamma$, equation (\ref{eq:Gamma}), and $\phi$,
equation (\ref{eq:dphidx}), we find that for a monotonic profile  
$F(\phi) = A|\theta(\phi) - \theta(0)|$. This result suggests 
that for a general profile we can also avoid performing the integration if 
we identify the zeros of $\Gamma$ and break apart the 
profile at those points so as to create a series of monotonic profiles. 
The absolute difference of the mixing angle across each monotonic region 
can be computed and the calculation for $F(\phi)$ is then an 
appropriate summation. The advantages of calculating $F(\phi)$ this way are: firstly, 
that it is far quicker than doing the integration, and secondly, $\Gamma$ can be somewhat 
noisy - as shown in figure (\ref{fig:Gammasolar}) - due to numerical problems 
associated with forming derivatives. 

To use the relationship between $\phi$ and $F$ one generates a 
pseudo-random number $u$ from a uniform probability 
distribution and sets $F=u$ before inserting this value into 
$\phi(F)$. There is one circumstance where inversion of $F(\phi)$ to $\phi(F)$ 
is not possible and this occurs whenever $|\Gamma| \propto|\theta'| 
=0$ over some extended distance within a profile. Such a 
region would possess a constant density. But over this region $S=1$ in the $\underline{b}$ basis 
so there is no need to perform the Monte Carlo calculation for this region. 
If this situation arises a simple solution is, again, to break apart the profile 
and only calculate the scattering matrix for those regions where $|\Gamma| \neq 0$.
In this way the total scattering matrix for the entire profile is the 
ordered product of the scattering matrices for each $|\Gamma| \neq 0$ zone. 

\subsection{Efficiently using the random $\phi$}

Once the $N_{S}$ values of $\phi$ have been found and stored in an array, a possible 
algorithm for $s$ would be:
\begin{enumerate}
\item use the first value, $\phi_{1}$, to calculate $s_{1}$  and add it to the unit matrix,
\item $\phi$-order the first two values, $\phi_{1}$ and $\phi_{2}$, calculate $s_{2}$, and add it to the $1+s_{1}$ sum,
\item repeat for all $N_{S}$ terms.
\end{enumerate}  
In this scheme each term in $s$ is calculated just once. 
But the presence of the weighting factors 
$w_{i}$ indicate that this is not optimal: we should calculate a term 
$s_{i}$ much more frequently if its weight is large and less frequently if 
the weight is small. There are many ways one can achieve a better 
load balancing: we adopted, after realizing that the labels on 
the $\phi$'s may be swapped amongst 
themselves, a scheme whereby we rewrite equation (\ref{eq:sappendix}) as 
\begin{widetext}
\begin{equation}
s = 1 + \sum_{j=1}^{N_{\phi}} 
\;\frac{(-\imath)}{_{N_{\phi}}C_{1}}\,h_{j} + \sum_{ \{j,k\} } 
\frac{(-\imath)^{2}}{2!\;_{N_{\phi}}C_{2}} 
\,\mathbb{T}(h_{j}\,h_{k}) + \ldots + 
\frac{(-\imath)^{N_{\phi}+1}}{(N_{\phi}+1)!}\,\mathbb{T}(h_{1}\,h_{2}\,h_{3}\,\ldots\,h_{N_{\phi}+1}) 
+ \ldots \label{eq:averaging}
\end{equation}
\end{widetext}
where $_{N_{\phi}}C_{i}$ are the binomial coefficients, 
$N_{\phi}$ is an integer that satisfies $1 \leq N_{\phi} \leq N_{S}$ and $\{ j,k\}$ 
indicates all combinations of two $\phi$'s from the first $N_{\phi}$ 
in the list. This equation expresses the fact that any element of 
the first $N_{\phi}$ values of $\phi$ from our array may be used 
to calculate $s_{1}$, any ordered pair of the first $N_{\phi}$ 
may be selected for $s_{2}$ and so up to $s_{N_{\phi}}$, 
thereafter we calculate the higher order terms as described 
before. The appearance of the binomial coefficients in the denominators 
has the effect of increasing the number of trials that will form the first 
$N_{\phi}$ terms of $S$. But the additional computation obviously leads to an increase in the 
amount of time required to generate just one $s$. To compensate 
for the longer runtime we can reduce the number of samples of $s$
that we average to form the scattering matrix. If $\tau(0)$ is the time 
required to form $s$ via equation (\ref{eq:sappendix}), and $\tau(N_{\phi})$ is the 
amount of time to calculate $s$ according to equation (\ref{eq:averaging}), then 
the number of $s$ samples that we would have averaged with equation (\ref{eq:sappendix}), 
which we call $N_{T}(0)$, is reduced to $N_{T}(N_{\phi})$ when we use 
equation (\ref{eq:averaging}) so as maintain $N_{T}(N_{\phi})\,\tau(N_{\phi}) = N_{T}(0)\,\tau(0)$. 
Even though the number of $s$ that we average to form the scattering matrix is reduced 
a judicious choice for $N_{\phi}$ and the presence of the binomial coefficients 
can more than compensate this loss so that our scattering matrix is more accurate and the 
code more efficient. We base our decision for selecting $N_{\phi}$ by defining a quantity 
$V_{S}$ as 
\begin{equation}
V_{S} = \left[ \frac{\sum_{i=1}^{N_{\phi}}\; w^{2}_{i} / 
_{N_{\phi}}C_{i} + \sum_{i=N_{\phi}+1}^{N_{S}}\,w^{2}_{i}}
             {  \sum_{i=1}^{N_{S}} w^{2}_{i} } \right]
             \; \left[ \frac{\tau(N_{\phi})}{\tau(0) } \right] \label{eq:Vs}
\end{equation}
and determine the value of $N_{\phi}$ that minimizes $V_{S}$.
The reader may find that an alternate selection criteria leads to 
a more efficient algorithm. 
The computation times $\tau$ were found by numerical experiments 
and the application of fitting formulae to the results although 
one may alternatively have some knowledge of their relative size 
from the design of the algorithm. 




\begin{thebibliography}{99}

\bibitem{M&S1986} S.P. Mikheev and A.I. Smirnov, Nuovo Cimento C, {\bf 9}, 17, (1986)

\bibitem{Wolfenstein1977} L. Wolfenstein, Phys. Rev., {\bf D17}, 2369 (1978)

\bibitem{SNO}
  Q.~R.~Ahmad {\it et al.}  [SNO Collaboration],
  Phys.\ Rev.\ Lett.\  {\bf 89}, 011301 (2002)
  [arXiv:nucl-ex/0204008].

\bibitem{Bethe1986} H.A. Bethe, Physical Review Letters, {\bf 56}, 1305 (1986)

\bibitem{Haxton1986} W.C. Haxton, Physical Review Letters, {\bf 57}, 1271 (1986)

\bibitem{mocioiu}
  I.~Mocioiu and R.~Shrock,
  Phys.\ Rev.\ D {\bf 62}, 053017 (2000)
  [arXiv:hep-ph/0002149].

\bibitem{fuller}
  G.~M.~Fuller, W.~C.~Haxton and G.~C.~McLaughlin,
  Phys.\ Rev.\ D {\bf 59}, 085005 (1999)
  [arXiv:astro-ph/9809164].

\bibitem{dighe}
  A.~S.~Dighe and A.~Y.~Smirnov,
  Phys.\ Rev.\ D {\bf 62}, 033007 (2000)
  [arXiv:hep-ph/9907423].

\bibitem{petzold81} L.R. Petzold, Siam J. Numer. Anal., {\bf 18}, 455 (1981)

\bibitem{SF2002} R.C. Schirato, and G.,M. Fuller, astro-ph/0205390

\bibitem{EMV2003} J. Engel, G.C. McLaughlin, and C. Volpe, Phys.Rev. {\bf D67}, 013005 (2003)

\bibitem{lunardini}
  C.~Lunardini and A.~Y.~Smirnov,
  JCAP {\bf 0306}, 009 (2003)
  [arXiv:hep-ph/0302033].

\bibitem{K&T2001} M. Kachelrie{\ss}, and R. Tom{\` a}s, \prd, {\bf 64}, 073002 (2001)

\bibitem{Friedland2001} A. Friedland, \prd, {\bf 64}, 013008 (2001)

\bibitem{Lillie67} H.W. Lilliefors, Journal of the ASA, {\bf 62}, 399 (1967)

\bibitem{Haxton1987} W.C. Haxton, \prd, {\bf 35}, 2352 (1987)

\bibitem{K&P1989} T.K. Kuo, and J. Pantaleone, Reviews of Modern Physics, {\bf 61}, 937 (1989)

\bibitem{BB1996} A.B. Balantekin, and J.F. Beacom, Phys.Rev., {\bf D54}, 6323 (1996)

\bibitem{LB1994} F.N. Loreti, and A.B. Balantekin, Phys.Rev. {\bf D50}, 4762 (1994)
\bibitem{Loretietal1995} F.N. Loreti, \emph{et al.}, \prd, {\bf 52}, 6664 (1995)
\bibitem{BFL1996} A.B. Balantekin, J.M. Fetter, and F.N. Loreti, Phys.Rev., {\bf D54}, 3941 (1996)
\bibitem{F&K2001} P.M. Fishbane, and P. Kaus, Journal of Physics G, {\bf 27}, 2405 (2001)
 
\bibitem{Bahcall2005} J.N. Bahcall, and A.M. Serenelli, and S. Basu, Ap. J. L., {\bf 621}, L85 (2005)

\bibitem{Heger} http://www.ucolick.org/~alex/stellarevolution/data.shtml

\bibitem{BBKM2005} J. Blondin, J. Brockman,, J.P. Kneller, and G.C. McLaughlin, in preparation 

\end{thebibliography}
\end{document}